\documentclass[sigconf]{acmart}

\AtBeginDocument{%
  }

\usepackage{algorithm}
\usepackage{algorithmic}
\usepackage{colortbl}
\usepackage{makecell}
\usepackage{multirow}
\usepackage{pifont}
\usepackage{soul}
\usepackage{subcaption}
\usepackage{balance}

\newcommand{\ours}{\textsc{AudioKV}}

\definecolor{mygray}{gray}{0.92}
\definecolor{ForestGreen}{RGB}{34,139,34}
\definecolor{ForestRed}{RGB}{220,50,50}
\definecolor{DarkPink}{RGB}{255,20,147}
\definecolor{DElectricBlue}{RGB}{93,117,131}
\definecolor{LightDElectricBlue}{RGB}{206,214,218}
\definecolor{LightGreen}{RGB}{220,245,220}

\copyrightyear{2026}
\acmYear{2026}
\setcopyright{cc}
\setcctype{by}

\acmConference[MM '26]
  {Proceedings of the 34th ACM International Conference on Multimedia}
  {November 10--14, 2026}
  {Rio de Janeiro, Brazil.}
\acmBooktitle{Proceedings of the 34th ACM International Conference on
  Multimedia (MM '26), November 10--14, 2026, Rio de Janeiro, Brazil}
\acmISBN{979-8-4007-2213-4/2026/11}

\acmDOI{10.1145/3767308.3835763}

\begin{document}

\title{\ours{}: KV Cache Eviction in Efficient Large Audio Language Models}

\author{Yuxuan Wang}
\authornote{Work conducted during an internship at Shanghai Jiao Tong University.}
\email{yxwang1215@gmail.com}
\affiliation{%
  \institution{Xidian University}
  \city{Xi'an}
  \country{China}}

\author{He Peize}
\email{2023300904027@std.uestc.edu.cn}
\affiliation{%
  \institution{University of Electronic Science and Technology of China}
  \city{Chengdu}
  \country{China}}

\author{Xiyan Gui}
\email{u202315217@hust.edu.cn}
\affiliation{%
  \institution{Huazhong University of Science and Technology}
  \city{Wuhan}
  \country{China}}

\author{Xiaoqian Liu}
\email{liuxiaoqian0319@outlook.com}
\affiliation{%
  \institution{Northeastern University}
  \city{Shenyang}
  \country{China}}

\author{Junhao He}
\email{hejunhao2024@sjtu.edu.cn}
\affiliation{%
  \institution{Shanghai Jiao Tong University}
  \city{Shanghai}
  \country{China}}

\author{Xuyang Liu}
\email{liuxuyang@stu.scu.edu.cn}
\affiliation{%
  \institution{Sichuan University}
  \city{Chengdu}
  \country{China}}

\author{Xuming Hu}
\authornote{Corresponding authors.}
\email{xuminghu97@gmail.com}
\affiliation{%
  \institution{The Hong Kong University of Science and Technology (Guangzhou)}
  \city{Guangzhou}
  \country{China}}

\author{Linfeng Zhang}
\authornotemark[2]
\email{zhanglinfeng1997@outlook.com}
\affiliation{%
  \institution{Shanghai Jiao Tong University}
  \city{Shanghai}
  \country{China}}
  
\renewcommand{\shortauthors}{Yuxuan Wang et al.}

\begin{abstract}
Large Audio-Language Models (LALMs) have set new benchmarks in speech
processing, yet their deployment is hindered by the memory footprint of the
Key-Value (KV) cache during long-context inference. While general KV cache
compression techniques excel in LLMs, they often fail in the audio domain by
overlooking the intrinsic temporal continuity of acoustic signals. To bridge
this gap, we propose AudioKV, a novel framework that robustly prioritizes
audio-critical attention heads through a hardware-friendly semantic-acoustic
alignment mechanism. Specifically, we identify these modality-specialized
heads by analyzing attention scores in ASR tasks and dynamically allocate KV
cache budgets preferentially to them. Furthermore, we introduce Spectral Score
Smoothing (SSS), an FFT-based global filtering strategy designed to suppress
high-frequency noise and recover smooth global trends from importance scores,
ensuring more balanced token selection with unprecedented precision. Extensive
evaluations across multiple LALMs, including the Qwen and Gemma series,
demonstrate that AudioKV significantly outperforms baselines while enhancing
computational efficiency. Notably, at a 40\% compression ratio, AudioKV
maintains near-full accuracy on Qwen3-Omni-30B with only a 0.45\% drop, whereas
traditional methods suffer from catastrophic performance degradation and
repetition. Our code is publicly available at
\url{https://github.com/yxwang1215/Audio_kvcache}.
\end{abstract}



 \maketitle

\section{Introduction}
\label{sec:intro}

Large Audio-Language Models (LALMs) with speech capabilities, such as Qwen2.5-Omni~\citep{xu2025qwen25omnitechnicalreport,xu2025qwen3omnitechnicalreport}, are increasingly deployed in real-world applications involving complex multimodal interactions. However, the Key-Value (KV) cache remains a critical bottleneck, as its memory footprint grows linearly with audio duration, severely limiting on-device and streaming deployment.

To effectively solve this problem, recent advances in KV cache compression for text-only LLMs show that significant memory savings are possible by retaining only salient past tokens. Methods such as SnapKV~\citep{NEURIPS2024_28ab4182} and AdaKV~\citep{feng2025adakvoptimizingkvcache} rely on attention-based importance estimation and perform well in text settings. Nevertheless, their direct applicability to the audio modality remains unclear due to the fundamentally different temporal structures and redundancy patterns, motivating us to systematically study the problems and corresponding efficient solutions of KV cache eviction in LALMs.

\begin{figure}[t]
    \centering
    \includegraphics[width=\linewidth]{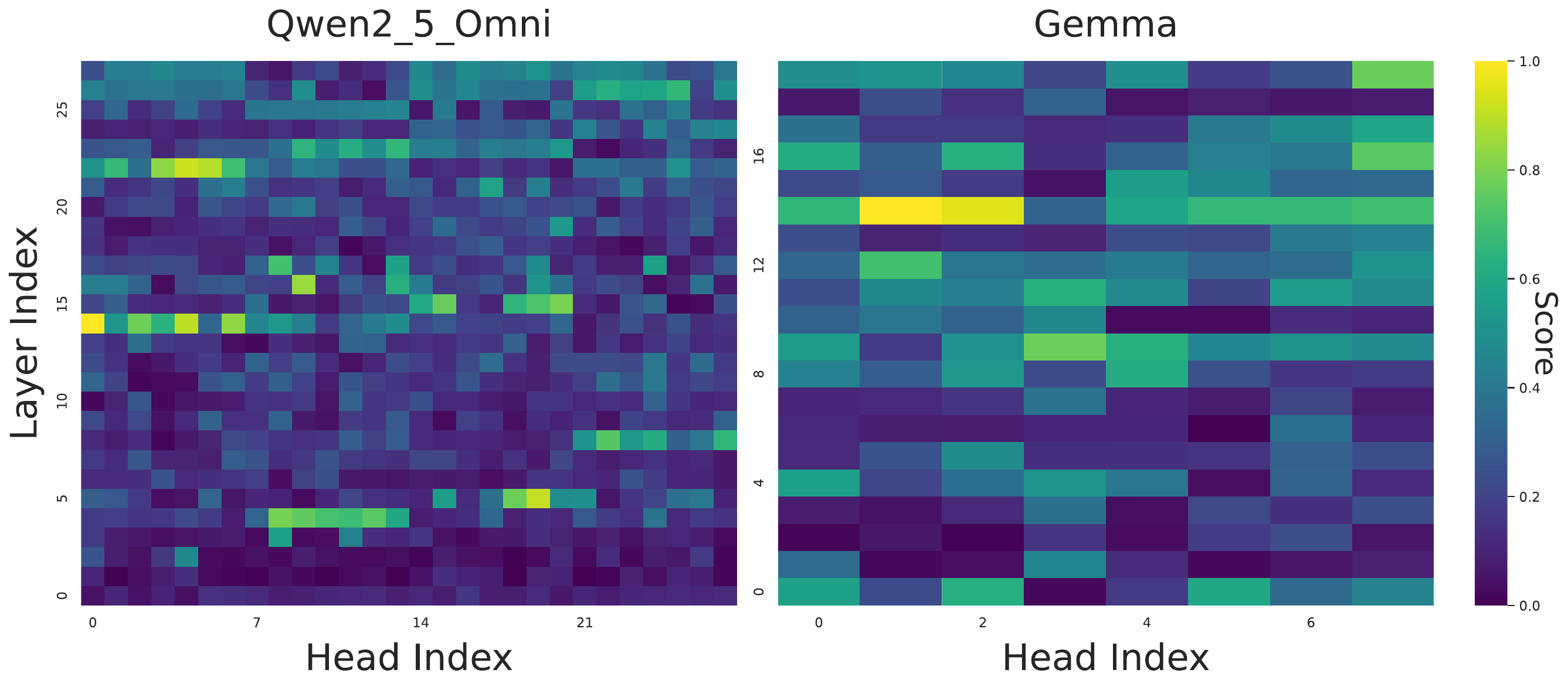}
    \caption{Visualization of audio critical heads in Qwen2.5-Omni-7B and Gemma-3n-E4B models.}
    \label{fig:head_salient}
\end{figure}

\paragraph{\textbf{Motivation 1: Modality-specialized attention heads call for audio-aware KV compression.}}
Prior studies show that attention heads in LALMs are modality-specialized~\citep{Wang_2025_ICCV}, suggesting that only a subset of attention heads captures acoustic information in audio-centric tasks, while others focus on linguistic or cross-modal patterns. This heterogeneity suggests that uniformly compressing the KV cache across all heads is suboptimal, as it may unnecessarily degrade audio-critical representations.

\paragraph{\textbf{Solution 1: Audio-aware KV cache allocation}}
We propose an audio-aware KV cache allocation strategy that exploits the unequal contribution of attention heads to acoustic modeling in LALMs. Specifically, we identify audio-critical heads by analyzing attention scores between paired audio tokens and decoded text tokens in an ASR task, and select heads that consistently exhibit higher audio–text attention. KV cache budgets are then allocated preferentially to these audio-critical heads, while stronger compression is applied to less relevant ones. This strategy preserves essential acoustic dependencies during decoding with substantially reduced memory overhead. \emph{As shown in Figure~\ref{fig:head_salient}, only a small subset of attention heads exhibits strong audio relevance}, indicating that acoustic modeling is concentrated in a limited number of heads and justifying the design of the proposed audio-aware KV cache allocation.

\begin{figure}[t]
    \centering
    \includegraphics[width=\linewidth]{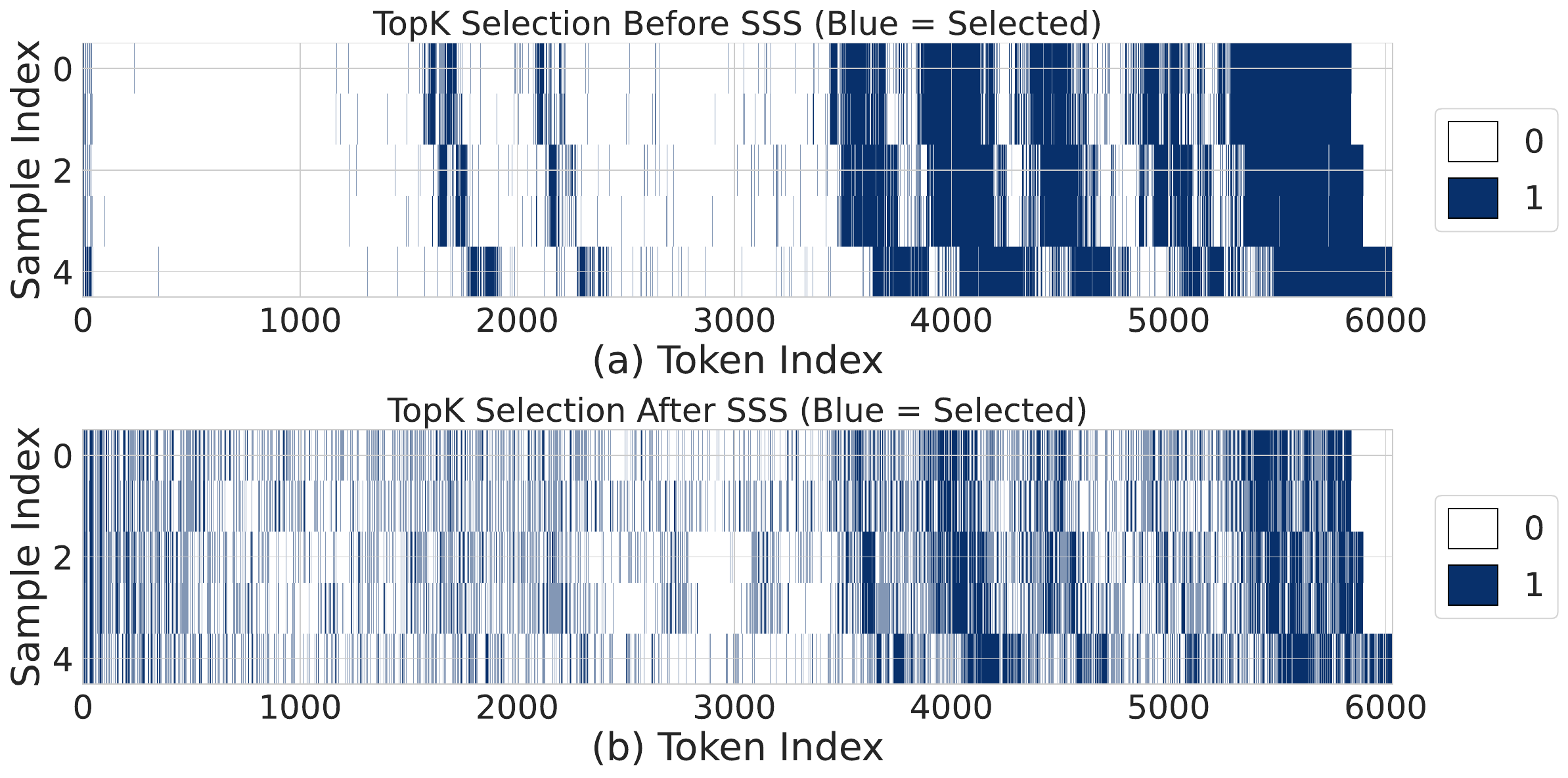}
    \caption{Distribution shift of Top-K token selection before and after spectral score smoothing (SSS).}
    \label{fig:SSS}
\end{figure}

\paragraph{\textbf{Motivation 2: Audio as a Smoothing and Continuous Signal.} }
Different from text, speech signals encode information in \textbf{temporally continuous acoustic representations}, where linguistic and semantic cues unfold gradually over time rather than being localized to isolated frames.
However, as illustrated in Figure~\ref{fig:SSS} (a), the original importance scores of audio tokens exhibit a strong temporal bias. Specifically, Top-K selection based on the raw importance scores tends to concentrate the preserved KV pairs within a small subset of token indices, forming dense clusters, while the remaining token indices are rarely selected, despite still carrying relevant contextual information. This uneven distribution violates the inherent temporal continuity of speech signals and leads to suboptimal KV cache utilization in practice.

\paragraph{\textbf{Solution 2: Spectral Score Smoothing (SSS)}}
To solve this problem, we introduce a signal-processing-based strategy to stabilize token-level importance estimation. Specifically, we treat the importance scores assigned by each attention head to past tokens as a one-dimensional temporal signal. In audio, such signals naturally exhibit multi-scale structure: slowly varying components capture stable semantic relevance, while rapid oscillations often reflect spurious local alignments.
Concretely, SSS applies an FFT-based global low-pass filter to suppress high-frequency noise and recover smooth global trends, and then interpolates the filtered signal with the original attention scores using a mixing coefficient~$\alpha$.
Unlike conventional pooling-based smoothing~\citep{NEURIPS2024_28ab4182}, which is inherently local and operates over a limited neighborhood, SSS performs smoothing over the entire importance signal, enabling global redistribution of attention scores.
As shown in Figure~\ref{fig:SSS}(b), 
this frequency-aware smoothing yields more balanced top-$K$ selections across the sequence, mitigating attention bias while remaining model-agnostic and computationally efficient.

Based on these key insights, we propose \textbf{AudioKV}, a novel framework for audio-centric KV cache compression. AudioKV integrates (i) head-aware KV cache allocation, which prioritizes memory for audio-critical attention heads, and (ii) Spectral Score Smoothing (SSS), which stabilizes token importance estimation by enforcing temporal continuity in acoustic signals and is compatible with existing score-based KV eviction methods. Extensive experiments on various ASR and ST benchmarks show that AudioKV achieves a superior accuracy–memory trade-off: at the same compression ratio, it substantially mitigates ASR degradation compared to head-agnostic baselines, and in many cases preserves near-full ASR accuracy while reducing KV memory to 40\% on Qwen3-Omni-30B-A3B-Instruct. These results demonstrate AudioKV as an effective and principled solution for efficient and scalable inference in LALMs.


In summary, our contributions are threefold:
\begin{itemize}
    \item We identify and quantify audio-critical attention heads in five large audio-language models, revealing strong head-level modality specialization in speech-centric decoding.
    \item We propose AudioKV, a head-aware KV cache compression framework that combines audio-head-aware budgeting with spectral score smoothing for audio tasks.
    \item Extensive empirical experiments demonstrate that AudioKV achieves superior accuracy under KV cache eviction scenarios while maintaining the same cache budget.
\end{itemize}

\vspace{-3mm}
\section{Related Work}
\label{sec:related}

\subsection{Attention and Speech--Text Grounding}
Attention-based ASR models, including LAS~\citep{7472621} and Transformer-based architectures~\citep{vaswani2017attention}, demonstrate that decoder attention implicitly encodes alignments between acoustic frames and output tokens. Large-scale systems such as Whisper~\citep{radford2023robust} and WhisperX~\citep{bain2023whisperx} explicitly provide word-level timestamps and confidence scores, which can serve as external alignment supervision. Existing analyses of Transformer attention heads in NLP and speech~\citep{DBLP:journals/corr/abs-1905-10650,DBLP:journals/corr/abs-1905-09418} primarily examine head behaviors at a coarse granularity, focusing on specialization or pruning effects, but they do not translate fine-grained grounding behavior into actionable inference-time policies. In contrast, our approach leverages word spans aligned by WhisperX together with top-K attention statistics to compute specific head-level audio-grounding scores, which are then effectively used to explicitly guide dynamic cache allocation strategies accurately.

\subsection{Efficient Inference under Long Context}
Prior work reduces memory usage for long-context inference through architectural sparsification, such as structured sparse attention and local attention windows~\citep{beltagy2020longformer}, as well as efficient attention implementations including FlashAttention~\citep{dao2022flashattention,dao2023flashattention,shah2024flashattention}. 
More recent KV-compression methods selectively retain past tokens based on token importance scores. Pioneering approaches like H$_2$O~\citep{NEURIPS2023_6ceefa7b} and StreamingLLM~\citep{ICLR2024_5e5fd18f} demonstrated that retaining only \textit{heavy-hitter} tokens or attention sinks can preserve generation quality. Building on this, methods such as SnapKV~\citep{NEURIPS2024_28ab4182} and AdaKV~\citep{feng2025adakvoptimizingkvcache} refine these selection metrics using local pooling heuristics, while PyramidKV~\citep{cai2025pyramidkvdynamickvcache} suggests allocating variable cache budgets across different layers. Parallel to pruning, quantization methods like KIVI~\citep{pmlr-v235-liu24bz} reduce memory footprints by quantizing stored key-value pairs.
SparseMM~\citep{Wang_2025_ICCV} exploits alignment signals with OCR-based supervision to learn sparse attention patterns, but primarily targets the visual domain. In contrast, our method focuses on the audio modality and introduces a \emph{head-wise} cache budgeting scheme that leverages grounding statistics to allocate larger global caches to audio-critical heads while reducing cache sizes for less informative ones.

\subsection{Spectral Smoothing for Sequence Importance}
Frequency-domain analysis has proven effective in sequence modeling and signal processing. 
FNet~\citep{DBLP:journals/corr/abs-2105-03824} and GFNet~\citep{rao2021globalfilternetworksimage} 
established that global spectral mixing can efficiently capture long-range dependencies, 
offering a competitive alternative to self-attention. A recent related 
work~\citep{lee2025frequencyawaretokenreductionefficient} has also applied frequency-domain 
analysis to accelerate inference; however, it targets token pruning in the visual domain, 
whereas our work focuses on KV cache compression for audio. Building upon the context of 
noisy attention scores observed in existing methods like SnapKV, which typically relies 
on pooling that is inherently local and sensitive to high-frequency noise. To address 
this, we propose Spectral Score Smoothing (SSS). Instead of simple pooling, SSS employs 
an RFFT-based global filter with an energy-driven cutoff across the frequency spectrum to isolate essential semantic features from semantic information, functioning 
as a plug-and-play module for multi-stage compression.
\section{Methodology}
\label{sec:method}

\begin{figure*}[!t] 
    \centering
    \includegraphics[width=\textwidth]{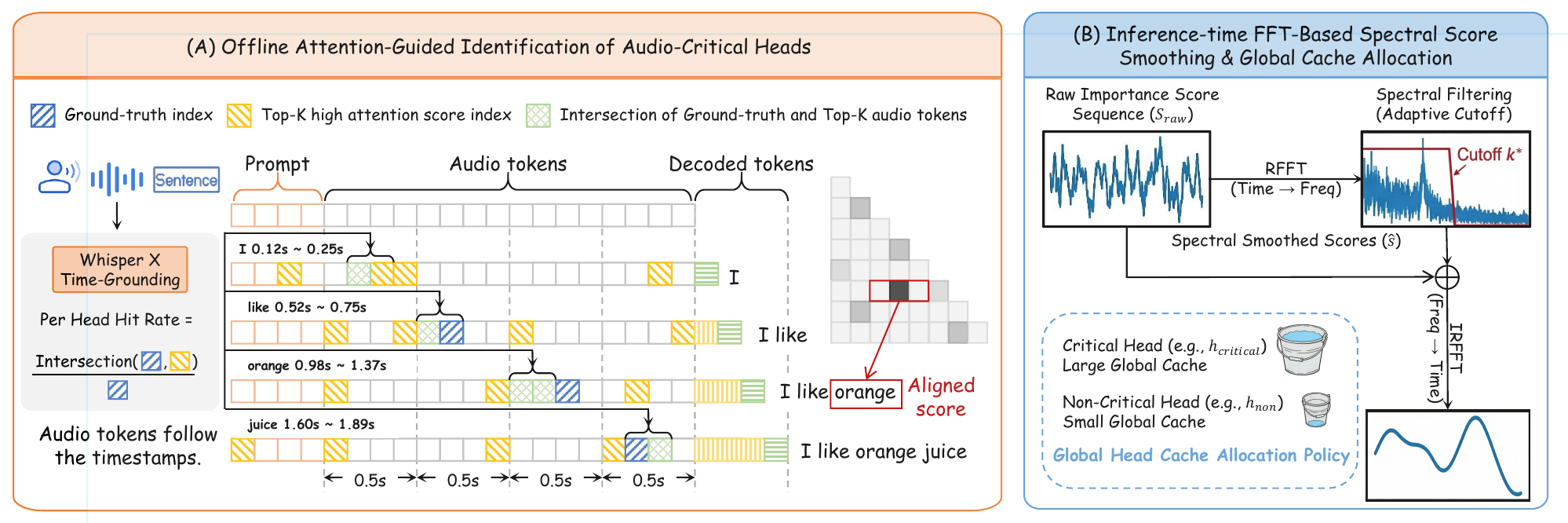}
    \caption{Overview of AudioKV.
(A) Offline identification of audio-critical attention heads. We analyze attention distributions to measure the overlap between high-attention audio tokens and ground-truth audio spans, enabling the selection of heads that are most relevant to audio modality.
(B) Inference-time lightweight plugin SSS, which performs spectral smoothing on importance scores to guide efficient cache allocation during decoding.}
    \label{fig:overview}
\end{figure*} 

\subsection{Preliminaries}

\paragraph{Modality-bias in Attention Heads.}
Multimodal Transformers with $L$ layers and $H$ attention heads exhibit strong functional heterogeneity: 
only a small fraction of heads consistently attends to non-textual tokens, while most heads primarily 
model textual context. This sparse and stable \emph{modality bias} has been observed across 
architectures and modalities. Prior work such as SparseMM~\citep{Wang_2025_ICCV} exploits this 
property to guide head-wise KV cache compression in the visual domain to improve overall 
inference efficiency.

\paragraph{Modality-aware Head-wise KV Cache Allocation.}
To adapt KV cache allocation to non-text modalities, we associate each attention head $(l,h)$ with a scalar \emph{modality score} $S_{l,h}$ that measures its relevance to a given modality. Let $\mathbf{S} \in \mathbb{R}^{L \times H}$ denote the collection of all head-wise scores. 

Given a total KV cache budget $B$, we allocate cache capacity asymmetrically across heads according to these scores. Specifically, a head $(l,h)$ is
\begin{equation}
b_{l,h} = w + r + B_{\text{score}} \cdot
\frac{S_{l,h}}{\sum_{l',h'} S_{l',h'}},
\end{equation}
where $w$ is a fixed local window size, $r$ is a uniform baseline allocation shared by all heads, and $B_{\text{score}}$ is the remaining budget distributed proportionally based on modality relevance. This modality-agnostic formulation provides a general foundation, which we instantiate concretely for the audio modality in later sections.

\subsection{Identifying Audio-Critical Heads}
\label{subsec:audio_head_selection}

Although SparseMM has demonstrated in the visual domain that retaining only a small fraction of visual heads suffices to significantly reduce the KV cache of redundant heads, it remains unclear whether attention heads exhibit similarly consistent patterns in the audio modality.

Our goal is to identify attention heads that are truly responsible for grounding ASR decoded tokens in the input audio tokens, and to leverage this information to drive a global, head-wise cache allocation policy. Drawing inspiration from SparseMM~\citep{Wang_2025_ICCV}, we align word-level ASR supervision with token-level attention behavior, explicitly leveraging the temporal structure of the input audio signal, and derive per-head importance scores from ASR tasks.

\paragraph{Word-level Audio Alignment.}
Given an input utterance, we obtain word-level timestamps and confidence scores from WhisperX.\footnote{\url{https://github.com/m-bain/whisperX}} Each word $w_i$ is represented as
\begin{equation}
    w_i = \big(\text{text}_i,\, t_i^{\text{start}},\, t_i^{\text{end}},\, s_i\big),
\end{equation}
where $t_i^{\text{start}}$ and $t_i^{\text{end}}$ denote the temporal boundaries of the word and $s_i$ is the alignment confidence. We retain only high-confidence words to treat them as reliable anchors.
\begin{equation}
\mathcal{W} = \{ w_i \mid s_i \ge \tau \}, \quad \tau = 0.95,
\end{equation}
We treat these identified words as reliable anchors to establish precise alignments between audio frames and textual outputs.

Let $\mathcal{A} = \{ a_0, a_1, \dots, a_{N_{\text{audio}}-1} \}$ denote the indices of audio tokens in the model input sequence, where $a_0$ is the starting index of the audio prefix. Assuming a uniform mapping between time and audio-token index, we convert each word’s timestamps into a contiguous span of audio token indices
\begin{equation}
\mathcal{A}(w_i) = \{ a_i^{\text{start}}, \dots, a_i^{\text{end}} \},
\end{equation}
with
\begin{equation}
\begin{aligned}
a_i^{\text{start}} &= a_0 + \left\lfloor \frac{t_i^{\text{start}}}{T} \cdot N_{\text{audio}} \right\rfloor, \\
a_i^{\text{end}}   &= a_0 + \left\lfloor \frac{t_i^{\text{end}}}{T} \cdot N_{\text{audio}} \right\rfloor .
\end{aligned}
\end{equation}

\paragraph{Aligning generated tokens to ASR words.}
During decoding, the model generates a sequence of text tokens $\{y_t\}_{t=1}^{T_{\text{gen}}}$. We decode these tokens into text and greedily align them to the high-quality word set $\mathcal{W}$. For each word $w_i$, we identify the corresponding set of decoding step indices
\begin{equation}
\mathcal{T}(w_i) \subseteq \{1, \dots, T_{\text{gen}}\},
\end{equation}
which represent the generation of that word.

\paragraph{Collecting Attention Hits to Audio Spans.}
We instrument the decoder with forward hooks on all transformer layers. For each decoding step $t$ and attention head $(\ell, h)$, we record the attention distribution from the newly generated token to all previous token indices
\begin{equation}
\mathbf{A}^{(\ell, h)}_t \in \mathbb{R}^{L_t},
\end{equation}
where $L_t$ is the current sequence length in indices. To focus on salient context tokens, we extract the top-$K$ attended token indices
\begin{equation}
\mathcal{S}^{(\ell, h)}_t = \operatorname{TopK}\big(\mathbf{A}^{(\ell, h)}_t, K\big),
\end{equation}
with typically set to 24 in experiments.

For a decoding step $t \in \mathcal{T}(w_i)$, we calculate the number of top-$K$ indices within the audio token span, as shown in Figure \ref{fig:overview}.

\begin{equation}
c^{(\ell, h)}_{t,\text{word}} =
\big|\mathcal{S}^{(\ell, h)}_t \cap \mathcal{A}(w_i)\big|.
\end{equation}
We then define a step-wise hit ratio
\begin{equation}
r^{(\ell, h)}_{t,\text{hit}} =
\frac{c^{(\ell, h)}_{t,\text{word}}}{K},
\end{equation}
which measures how frequently an attention head’s most salient attention score is allocated to the correct audio token indices when generating a word.

\paragraph{Head Importance Estimation.}
Aggregating over all word-aligned decoding step indices $\mathcal{T}_{\text{ASR}}$, we define a global importance score for each head:
\begin{equation}
S^{(\ell, h)}_{\text{hit}} =
\mathbb{E}_{t \in \mathcal{T}_{\text{ASR}}}
\big[ r^{(\ell, h)}_{t,\text{hit}} \big].
\end{equation}

Heads with high $S^{(\ell, h)}_{\text{hit}}$ consistently allocate their most salient attention to the audio token indices corresponding to the words being generated, indicating strong audio grounding behavior.

\paragraph{Head-wise Cache Allocation.}
At inference time, we use $S^{(\ell, h)}_{\text{hit}}$ as a prior over attention heads to guide global cache allocation. Given a total cache budget $B$, each head is assigned a cache capacity
\begin{equation}
B^{(\ell, h)} =
\left\lfloor
B \cdot
\frac{S^{(\ell, h)}_{\text{hit}}}
{\sum_{\ell',h'} S^{(\ell',h')}_{\text{hit}}}
\right\rfloor.
\end{equation}
This allocation strategy prioritizes heads that are empirically responsible for aligning generated words with their corresponding audio token spans, while reducing cache usage for non-audio heads.

\begin{table*}[!t]
\centering
\vspace{-2mm}
\caption{Main results of \textbf{AudioKV} across various KV-cache compression budgets on ASR and ST benchmarks. For each dataset and model, the best-performing method is highlighted in \textbf{bold}. Numbers in the header row (\emph{e.g., 0.8, 0.6}) denote the KV-cache \textbf{retention ratio}. 
\textbf{Note:} When models exhibit degenerate repetition, insertion errors escalate significantly, leading to high WER. For readability, we report ASR accuracy as
$\max(100-\mathrm{ER}, 0)$, where ER denotes CER or WER.
Entries shown as 0 indicate failure to generate meaningful output.}
\label{tab:main_result}
\vspace{-2mm}

\resizebox{\textwidth}{!}{%
\tiny
\setlength{\tabcolsep}{1.5pt}
\renewcommand{\arraystretch}{1.00}
\begin{tabular}{l *{21}{c} c}
\toprule
\multirow{2}{*}{\textbf{Methods}} &
\multicolumn{15}{c}{\textbf{Automatic Speech Recognition (ASR)}} &
\multicolumn{6}{c}{\textbf{Speech Translation (ST)}} &
\multirow{2}{*}{\textbf{Avg.}} \\
\cmidrule(lr){2-16} \cmidrule(lr){17-22}
& \multicolumn{3}{c}{ZH} & \multicolumn{3}{c}{EN} & \multicolumn{3}{c}{FR} & \multicolumn{3}{c}{DE} & \multicolumn{3}{c}{ES}
& \multicolumn{3}{c}{E2C} & \multicolumn{3}{c}{C2E}
& \\
& 0.8 & 0.6 & 0.4 & 0.8 & 0.6 & 0.4 & 0.8 & 0.6 & 0.4 & 0.8 & 0.6 & 0.4 & 0.8 & 0.6 & 0.4
& 0.8 & 0.6 & 0.4 & 0.8 & 0.6 & 0.4 & \\
\midrule

\noalign{\vskip -1.8pt}
\multicolumn{23}{c}{\textbf{Qwen2.5-Omni-3B}} \\
\noalign{\vskip -1.4pt}
\cmidrule(lr){1-23}
\textcolor{gray}{Full KV}
& \multicolumn{3}{c}{91.8}
& \multicolumn{3}{c}{94.7}
& \multicolumn{3}{c}{90.9}
& \multicolumn{3}{c}{88.9}
& \multicolumn{3}{c}{94.0}
& \multicolumn{3}{c}{35.4}
& \multicolumn{3}{c}{25.2}
& 74.4 \\
SnapKV
& 0.0 & 0.0 & 0.0
& 67.0 & 15.7 & 0.0
& 54.2 & 0.0 & 0.0
& 0.0 & 0.0 & 0.0
& 45.0 & 0.0 & 0.0
& 34.6 & 31.5 & 24.5
& 24.2 & 21.2 & 15.1
& 15.9 \\
AdaKV
& 0.0 & 0.0 & 0.0
& 65.0 & 11.5 & 0.0
& 37.1 & 0.0 & 0.0
& 0.0 & 0.0 & 0.0
& 22.4 & 0.0 & 0.0
& 34.6 & 31.9 & 25.2
& 24.4 & 21.4 & 15.8
& 13.8 \\
PyramidKV
& 0.0 & 0.0 & 0.0
& 24.3 & 0.0 & 0.0
& 0.0 & 0.0 & 0.0
& 0.0 & 0.0 & 0.0
& 0.0 & 0.0 & 0.0
& 34.1 & 27.3 & 22.6
& 23.7 & 17.0 & 13.8
& 7.8 \\
H2O
& 0.0 & 0.0 & 0.0
& 39.9 & 0.0 & 0.0
& 0.0 & 0.0 & 0.0
& 0.0 & 0.0 & 0.0
& 0.0 & 0.0 & 0.0
& 30.3 & 23.8 & 14.9
& 20.0 & 14.6 & 9.1
& 7.3 \\
 \textbf{AudioKV}
& \textbf{91.7} & \textbf{90.8} & 0.0
& \textbf{95.0} & \textbf{93.9} & \textbf{90.0}
& \textbf{89.9} & \textbf{90.5} & \textbf{85.8}
& \textbf{89.5} & \textbf{90.2} & \textbf{75.8}
& \textbf{93.9} & \textbf{93.0} & \textbf{91.3}
& \textbf{35.3} & \textbf{35.0} & \textbf{32.5}
& \textbf{25.2} & \textbf{25.1} & \textbf{23.2}
& \textbf{68.5} \\
\midrule

\noalign{\vskip -1.8pt}
\multicolumn{23}{c}{\textbf{Qwen2.5-Omni-7B}} \\
\noalign{\vskip -1.4pt}
\cmidrule(lr){1-23}
\textcolor{gray}{Full KV}
& \multicolumn{3}{c}{93.5}
& \multicolumn{3}{c}{98.1}
& \multicolumn{3}{c}{83.9}
& \multicolumn{3}{c}{90.4}
& \multicolumn{3}{c}{94.5}
& \multicolumn{3}{c}{35.8}
& \multicolumn{3}{c}{25.6}
& 74.5 \\
SnapKV
& 61.1 & 0.0 & 0.0
& 80.7 & 40.2 & 0.0
& 72.4 & 0.0 & 0.0
& 64.6 & 0.0 & 0.0
& 76.0 & 0.0 & 0.0
& 35.3 & 33.6 & 27.5
& 25.2 & 23.4 & 18.7
& 26.6 \\
AdaKV
& 15.8 & 0.0 & 0.0
& 91.3 & 56.0 & 0.0
& 70.0 & 0.0 & 0.0
& 61.0 & 0.0 & 0.0
& 78.7 & 0.0 & 0.0
& 35.2 & 33.6 & 28.4
& 25.2 & 23.4 & 19.3
& 25.6 \\
PyramidKV
& 0.0 & 0.0 & 0.0
& 29.5 & 0.0 & 0.0
& 14.7 & 0.0 & 0.0
& 0.0 & 0.0 & 0.0
& 5.3 & 0.0 & 0.0
& 35.3 & 30.0 & 25.1
& \textbf{26.0} & 20.3 & 13.8
& 9.5 \\
H2O
& 0.0 & 0.0 & 0.0
& 64.0 & 6.4 & 0.0
& 31.1 & 0.0 & 0.0
& 5.3 & 0.0 & 0.0
& 14.9 & 0.0 & 0.0
& 32.7 & 27.6 & 17.8
& 22.8 & 18.6 & 12.2
& 12.1 \\
 \textbf{AudioKV}
& \textbf{93.1} & \textbf{93.1} & 0.0
& \textbf{98.1} & \textbf{98.0} & \textbf{89.1}
& \textbf{83.4} & \textbf{82.9} & \textbf{78.7}
& \textbf{90.3} & \textbf{90.8} & \textbf{85.3}
& \textbf{94.5} & \textbf{94.2} & \textbf{91.2}
& \textbf{35.7} & \textbf{35.2} & \textbf{32.8}
& 25.6 & \textbf{25.5} & \textbf{23.5}
& \textbf{68.6} \\
\midrule

\noalign{\vskip -1.8pt}
\multicolumn{23}{c}{\textbf{Qwen3-Omni-30B-A3B-Instruct}} \\
\noalign{\vskip -1.4pt}
\cmidrule(lr){1-23}
\textcolor{gray}{Full KV}
& \multicolumn{3}{c}{93.7}
& \multicolumn{3}{c}{98.3}
& \multicolumn{3}{c}{95.6}
& \multicolumn{3}{c}{95.5}
& \multicolumn{3}{c}{96.9}
& \multicolumn{3}{c}{40.1}
& \multicolumn{3}{c}{26.9}
& 78.1 \\
SnapKV
& 0.0 & 0.0 & 0.0
& 63.6 & 45.2 & 0.0
& 68.7 & 0.0 & 0.0
& 58.0 & 0.0 & 0.0
& 68.2 & 0.0 & 0.0
& 38.9 & 35.0 & 25.6
& 25.6 & 22.4 & 15.3
& 22.2 \\
AdaKV
& \textbf{93.6} & 0.0 & 0.0
& 81.9 & 75.6 & 60.6
& \textbf{95.6} & 90.9 & 0.0
& 94.8 & 0.0 & 0.0
& 96.2 & 84.8 & 0.0
& \textbf{40.1} & \textbf{40.1} & 24.1
& 26.3 & 24.3 & 18.9
& 45.1 \\
PyramidKV
& 56.8 & 24.2 & 0.0
& 66.1 & 0.0 & 0.0
& 33.9 & 0.0 & 0.0
& 0.6 & 0.0 & 0.0
& 27.4 & 0.0 & 0.0
& 39.0 & 34.4 & 29.9
& 25.2 & 23.6 & 10.7
& 17.7 \\
H2O
& 0.0 & 0.0 & 0.0
& 82.4 & 39.9 & 0.0
& 0.0 & 0.0 & 0.0
& 94.9 & 89.2 & 0.0
& 0.0 & 0.0 & 0.0
& 34.8 & 31.3 & 22.3
& 21.7 & 19.4 & 19.0
& 21.7 \\
 \textbf{AudioKV}
& 93.5 & \textbf{93.0} & 0.0
& \textbf{98.2} & \textbf{98.2} & \textbf{97.8}
& \textbf{95.6} & \textbf{95.5} & \textbf{89.2}
& \textbf{95.5} & \textbf{95.5} & \textbf{86.0}
& \textbf{96.9} & \textbf{96.8} & \textbf{83.6}
& 40.0 & 39.1 & \textbf{32.1}
& \textbf{26.8} & \textbf{26.2} & \textbf{20.6}
& \textbf{71.4} \\
\midrule

\noalign{\vskip -1.8pt}
\multicolumn{23}{c}{\textbf{Gemma-3n-E2B}} \\
\noalign{\vskip -1.4pt}
\cmidrule(lr){1-23}
\textcolor{gray}{Full KV}
& \multicolumn{3}{c}{35.6}
& \multicolumn{3}{c}{90.2}
& \multicolumn{3}{c}{57.2}
& \multicolumn{3}{c}{77.6}
& \multicolumn{3}{c}{76.9}
& \multicolumn{3}{c}{22.8}
& \multicolumn{3}{c}{12.2}
& 53.2 \\
SnapKV
& 35.1 & 32.5 & 3.7
& 0.0 & 0.0 & 0.0
& 33.8 & 0.0 & 0.0
& 54.3 & 45.0 & 0.0
& 74.0 & 47.3 & 0.0
& 19.2 & 15.2 & 5.0
& 12.1 & 10.2 & 1.6
& 18.5 \\
AdaKV
& 35.0 & 33.7 & 13.0
& 82.3 & 56.9 & 0.0
& 49.4 & 31.2 & 4.1
& 73.1 & 55.7 & 0.0
& 79.0 & 66.3 & 0.0
& 22.9 & 20.9 & 13.4
& 12.2 & 11.0 & 4.6
& 31.7 \\
PyramidKV
& 35.0 & 30.2 & 14.7
& 90.0 & 61.9 & 0.0
& 50.3 & 35.4 & 5.1
& 75.1 & 66.7 & 16.7
& 79.7 & 68.9 & 6.5
& 23.3 & 18.5 & 11.4
& 11.8 & 9.1 & 3.2
& 34.0 \\
H2O
& 34.5 & 33.6 & 11.5
& 83.0 & 41.2 & 0.0
& 47.2 & 25.3 & 0.0
& 71.5 & 50.5 & 0.0
& 68.9 & 36.3 & 0.0
& 22.8 & 20.1 & 12.3
& 12.2 & 10.8 & 4.0
& 27.9 \\
 \textbf{AudioKV}
& \textbf{36.3} & \textbf{34.6} & \textbf{31.2}
& \textbf{90.2} & \textbf{88.5} & \textbf{60.6}
& \textbf{66.8} & \textbf{64.9} & \textbf{28.9}
& \textbf{79.3} & \textbf{75.9} & \textbf{45.9}
& \textbf{82.9} & \textbf{81.2} & \textbf{36.2}
& \textbf{24.8} & \textbf{23.1} & \textbf{18.9}
& \textbf{12.3} & \textbf{11.7} & \textbf{9.0}
& \textbf{47.8} \\
\midrule

\noalign{\vskip -1.8pt}
\multicolumn{23}{c}{\textbf{Gemma-3n-E4B}} \\
\noalign{\vskip -1.4pt}
\cmidrule(lr){1-23}
\textcolor{gray}{Full KV}
& \multicolumn{3}{c}{47.3}
& \multicolumn{3}{c}{92.6}
& \multicolumn{3}{c}{60.5}
& \multicolumn{3}{c}{72.7}
& \multicolumn{3}{c}{80.6}
& \multicolumn{3}{c}{29.6}
& \multicolumn{3}{c}{14.5}
& 56.8 \\
SnapKV
& 44.4 & 40.6 & 0.0
& 0.0 & 0.0 & 0.0
& 57.8 & 0.0 & 0.0
& 8.5 & 0.0 & 0.0
& 73.1 & 39.2 & 0.0
& 21.7 & 17.4 & 5.7
& 14.0 & 12.0 & 2.3
& 16.0 \\
AdaKV
& 45.0 & 43.9 & 0.0
& 84.7 & 49.7 & 0.0
& 71.7 & 62.1 & 0.0
& 68.7 & 53.2 & 0.0
& 83.1 & 72.5 & 0.0
& 28.1 & 26.4 & 13.4
& 14.4 & 13.2 & 7.1
& 35.1 \\
PyramidKV
& 45.3 & \textbf{44.5} & 30.3
& 89.3 & 75.2 & 3.4
& 72.8 & 68.6 & 0.2
& 70.8 & 62.7 & 34.0
& 84.3 & 79.4 & 31.9
& 28.3 & 26.0 & 18.5
& 14.1 & 13.2 & 7.6
& 42.9 \\
H2O
& 44.6 & 43.4 & 0.0
& 89.8 & 54.6 & 0.0
& 51.6 & 36.0 & 0.0
& 66.9 & 49.8 & 0.0
& 74.5 & 51.9 & 0.0
& 28.2 & 25.7 & 19.6
& 14.0 & 12.9 & 6.2
& 31.9 \\
 \textbf{AudioKV}
& \textbf{45.4} & \textbf{44.5} & \textbf{39.3}
& \textbf{92.6} & \textbf{91.7} & \textbf{80.8}
& \textbf{75.5} & \textbf{75.2} & \textbf{35.5}
& \textbf{72.5} & \textbf{71.7} & \textbf{55.4}
& \textbf{86.6} & \textbf{86.4} & \textbf{61.1}
& \textbf{29.7} & \textbf{29.2} & \textbf{24.8}
& \textbf{14.5} & \textbf{14.4} & \textbf{12.2}
& \textbf{54.2} \\
\bottomrule
\end{tabular}%
}

\vspace{-2mm}
\end{table*}

\subsection{FFT-Based Spectral Attention Smoothing}
\label{sec:fft-smoothing}
\paragraph{\textbf{S}pectral \textbf{S}core \textbf{S}moothing for Cache Eviction.}
To enhance the robustness of key-value (KV) cache selection and reduce
high-frequency noise in attention trajectories, we introduce
\emph{Spectral Score Smoothing} (SSS), an FFT-based global filtering
module on the per-token attention importance scores.
Unlike pooling smoothing adopted by SnapKV~\citep{NEURIPS2024_28ab4182}, which is inherently local, frequency-domain filtering provides a global view over the entire score sequence and enables explicit control over frequency components.

Let $\mathbf{s} \in \mathbb{R}^{L}$ denote the 
per-head attention importance sequence
(for example, the averaged attention over a sliding decoding window).
We first apply a real-valued fast Fourier transform:
\begin{equation}
    \mathbf{F} = \mathrm{RFFT}(\mathbf{s}) \in \mathbb{C}^{L/2+1}.
\end{equation}
We decompose $\mathbf{F}$ into magnitude and phase components:
\begin{equation}
    \mathbf{m} = |\mathbf{F}|, \qquad 
    \phi = \angle \mathbf{F}.
\end{equation}

\paragraph{Adaptive Energy-Based Cutoff.}
To determine the spectral cutoff, we compute the cumulative energy of the
frequency magnitudes:
\begin{equation}
    E(k) = \sum_{i=0}^{k} m_i^{2},
\end{equation}
and select the smallest index $k^{\ast}$ satisfying
\begin{equation}
    E(k^{\ast}) \ge \rho \, E(L/2),
\end{equation}
where $\rho \in (0,1)$ is a user-configurable cutoff ratio.
This energy-driven cutoff automatically adapts to the smoothness of the
score distribution, retaining the dominant low-frequency components
while suppressing high-frequency noise.

\paragraph{Spectral Filtering.}
Based on the selected cutoff frequency $k^{\ast}$, we apply a low-pass spectral filter by constructing frequency mask $w_i$:
\begin{equation}
  w_i =
  \left\{
  \begin{array}{ll}
    1, & i \le k^{\ast}, \\
    0, & i > k^{\ast}.
  \end{array}
  \right.
\end{equation}
To reduce ringing artifacts caused by sharp frequency truncation,
we further apply a short cosine-shaped transition band around $k^{\ast}$,
which smoothly attenuates frequencies near the cutoff.

The filtered spectrum is then given by
\begin{equation}
    \tilde{\mathbf{F}} = \mathbf{F} \odot \mathbf{w},
\end{equation}
where $\odot$ denotes element-wise multiplication.

\paragraph{Reconstruction and Residual Mixing.}
We transform the filtered spectrum back into the time domain via an inverse FFT:
\begin{equation}
    \tilde{\mathbf{s}} = \mathrm{IRFFT}(\tilde{\mathbf{F}}, L).
\end{equation}
To preserve local attention structure and avoid smoothing, the final importance score is obtained through residual interpolation.

\begin{equation}
    \hat{\mathbf{s}} = (1 - \alpha)\,\mathbf{s}
      + \alpha\,\tilde{\mathbf{s}},
\end{equation}
where $\alpha \in [0,1]$ controls the strength of spectral smoothing.



\begin{figure*}[t]
    \centering 
    \includegraphics[width=\textwidth]{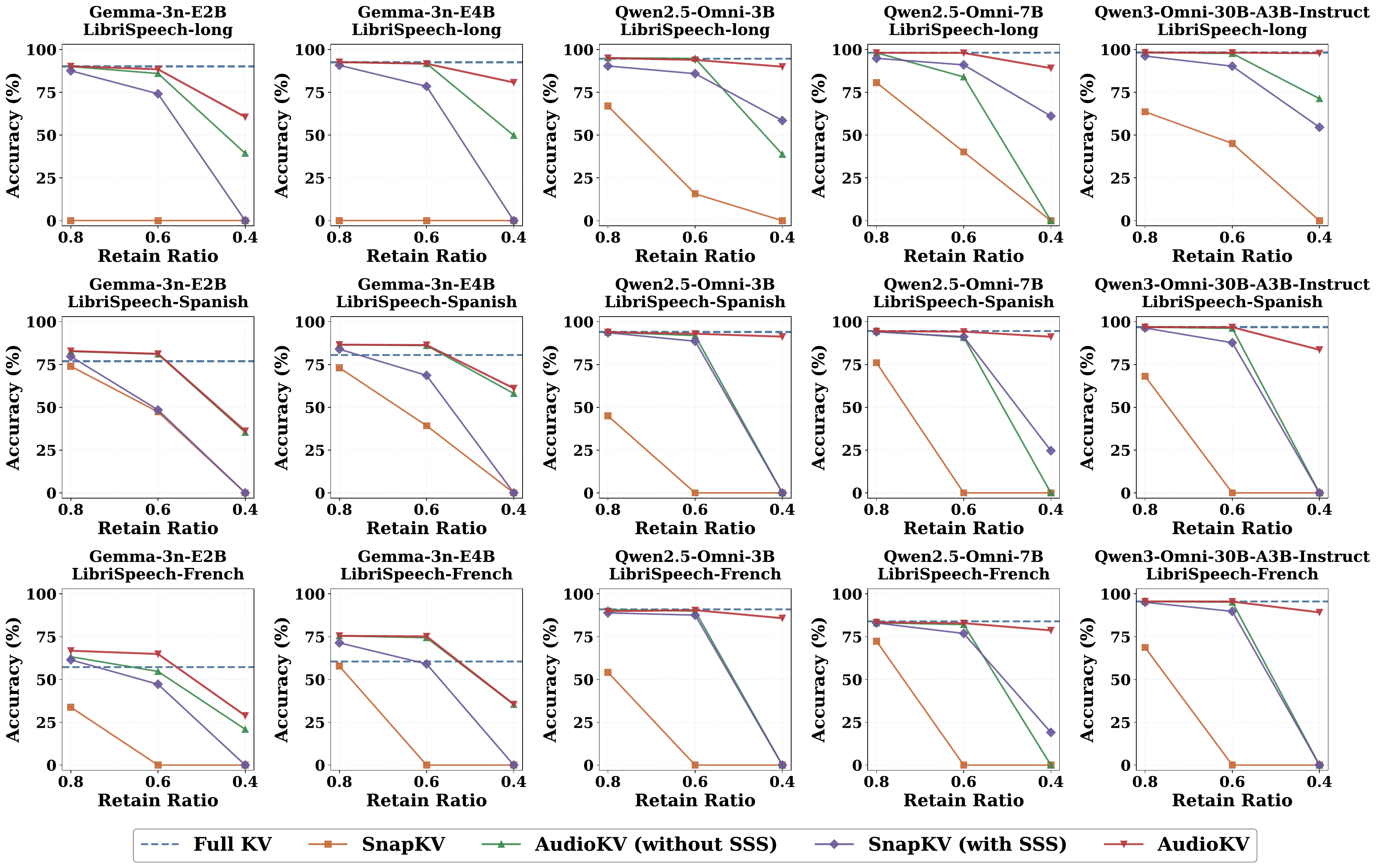}
    \caption{Performance comparison of KV cache methods across five models with varying retain ratios on LibriSpeech-long and Multilingual LibriSpeech datasets.}
    \label{fig:ablation_study}
\end{figure*}

\section{Experiments}
\label{sec:experiments}





\subsection{Experimental Setup}

\paragraph{Datasets and Benchmarks.}
To evaluate the effectiveness and generalizability of \textbf{AudioKV}, we conduct experiments on official test splits spanning \textbf{Automatic Speech Recognition (ASR)}, \textbf{Speech Translation (ST)}, and \textbf{Audio Question Answering (AQA)}. For ASR, we use \textbf{KeSpeech}~\citep{Tang2021KeSpeechAO} for Mandarin speech, \textbf{LibriSpeech-long (clean)}~\citep{xu2025qwen25omnitechnicalreport,xu2025qwen3omnitechnicalreport} for long-form English speech, and French, German, and Spanish subsets of \textbf{Multilingual LibriSpeech (MLS)}~\citep{Pratap2020MLSAL} for multilingual evaluation. For ST, we evaluate English-to-Chinese and Chinese-to-English translation on \textbf{CoVoST2}~\citep{wang21s_interspeech}. For AQA, we use \textbf{UltraEval-Audio} benchmarks~\citep{shi2026ultraeval}: \textbf{speech-chatbot-alpaca-eval}, \textbf{speech-web-questions}, \textbf{llama-questions}, and \textbf{speech-triviaqa}.

\paragraph{Models and Baselines.}
We validate AudioKV on a diverse set of Large Audio-Language Models (LALMs) spanning different architectures and parameter scales. These include the \textbf{Gemma-3n} family, with \textbf{Gemma-3n-E2B} and \textbf{Gemma-3n-E4B}~\citep{gemma_3n_2025}, and the \textbf{Qwen-Omni} family, with \textbf{Qwen2.5-Omni-3B}, \textbf{Qwen2.5-Omni-7B}, and \textbf{Qwen3-Omni-30B-A3B-Instruct}~\citep{xu2025qwen25omnitechnicalreport,xu2025qwen3omnitechnicalreport}. We compare AudioKV with the uncompressed \textbf{Full KV} setting and four representative KV cache compression methods: \textbf{SnapKV}~\citep{NEURIPS2024_28ab4182}, \textbf{AdaKV}~\citep{feng2025adakvoptimizingkvcache}, \textbf{PyramidKV}~\citep{cai2025pyramidkvdynamickvcache}, and \textbf{H2O}~\citep{NEURIPS2023_6ceefa7b}.

\paragraph{Evaluation Metrics.}
We use standard task-specific metrics for evaluation. For ASR, we report accuracy as $100-\text{Error Rate}$. Specifically, Character Error Rate (\textbf{CER}) is used for the Mandarin KeSpeech benchmark, while Word Error Rate (\textbf{WER}) is used for LibriSpeech and MLS. For ST, we report \textbf{chrF} for En-to-Zh translation, which better captures character-level correspondence in Chinese, and \textbf{BLEU} for Zh-to-En translation based on modified $n$-gram precision. For AQA, we adopt an \textbf{LLM-as-a-Judge} protocol~\citep{gu2025surveyllmasajudge}, where \textbf{Gemini-2.5-Flash}~\citep{comanici2025gemini25pushingfrontier} compares each generated answer with its reference answer and assigns a binary correctness score. We report the resulting judgment accuracy.

\subsection{Main Results}

\label{sec:main_results}
\paragraph{Overall Performance.}
As shown in Table~\ref{tab:main_result}, AudioKV consistently achieves lower Word Error Rate (WER) than the baselines across compression budgets. Its advantage is particularly pronounced under aggressive compression (e.g., budget = 0.4), where competing methods often suffer repetition and generation degeneration, producing WER values above 100 on long-form speech. In contrast, AudioKV maintains coherent transcriptions on \textit{LibriSpeech-long}. On \textit{Qwen3-Omni-30B-A3B-Instruct}, it incurs only a \textbf{0.45\% absolute accuracy drop} at \textbf{40\% KV retention}, demonstrating a strong efficiency--accuracy trade-off.

\paragraph{ASR vs. Speech Translation.}
AudioKV yields larger gains on ASR than on Speech Translation. Under aggressive compression, competing baselines are more prone to repetitive generation on long-form ASR, sharply increasing WER and potentially making $100-\mathrm{WER}$ negative before clipping. AudioKV substantially mitigates this failure mode. In comparison, baseline degradation is less severe on Speech Translation, resulting in smaller gains.

\paragraph{Stability Across Models and Datasets.}
AudioKV remains stable across datasets and architectures. Although \textit{PyramidKV} slightly outperforms AudioKV on \textit{LibriSpeech-long} with \textit{Gemma-3n-E2B} at budget = 0.8, it degrades sharply at budget = 0.4, highlighting AudioKV's robustness under aggressive compression.

\subsection{Ablation Studies}

\paragraph{Audio-Head-Aware Cache Allocation.}
We evaluate the effectiveness of audio-head-wise cache allocation by comparing SnapKV and AudioKV, both with and without SSS. While SnapKV allocates cache uniformly, AudioKV (w/o SSS) degenerates into SnapKV only if all heads are assigned equal importance. As shown in Figure~\ref{fig:ablation_study}, AudioKV (w/o SSS) consistently outperforms SnapKV, demonstrating the advantage of head-specific allocation. Furthermore, AudioKV’s superiority over SnapKV (w/ SSS) confirms that head-aware budget distribution is a significant contributor to the overall performance.

\paragraph{Spectral Score Smoothing (SSS)}
To demonstrate the significance of Spectral Score Smoothing across \textbf{SnapKV-based} and \textbf{AudioKV-based} settings, we assess the impact of SSS by comparing (i) SnapKV vs.\ SnapKV (w/ SSS) and (ii) AudioKV (w/o SSS) vs.\ AudioKV. As reported in Figure~\ref{fig:ablation_study}, SSS consistently yields substantial performance gains under both uniform and head-wise cache allocation. Notably, SnapKV (w/ SSS) with a cache budget of 0.6 surpasses vanilla SnapKV with a larger budget of 0.8, highlighting the efficiency benefits of SSS. Similar improvements are observed for AudioKV, indicating that SSS robustly enhances cache utilization by stabilizing spectral attention.

\begin{table}[ht]
\centering
\caption{Comparison of KV cache compression methods on AQA benchmarks. All experiments use a 40\% KV cache budget, max\_gen\_tokens=64, and Gemini 2.5 Flash as the judge. Bold indicates the best performance.}
\label{tab:aqa-generalization}

\centering 
\small 
\setlength{\tabcolsep}{6pt} 
\begin{tabular}{l cccc} 
\toprule
\textbf{Method} & \textbf{S-Alpaca} & \textbf{Llama-Q} & \textbf{S-Trivia} & \textbf{S-Web} \\
\midrule
\rowcolor[gray]{0.95} \multicolumn{5}{c}{\textbf{Qwen2.5-Omni-3B}} \\
SnapKV    & 54.5 & 72.0 & 27.4 & 44.9 \\
AdaKV      & 56.1 & 74.7 & 30.6 & 44.5 \\
PyramidKV & 56.1 & 74.7 & 28.3 & 45.3 \\
AudioKV    & \textbf{57.6} & \textbf{75.0} & \textbf{31.9} & \textbf{46.1} \\
\midrule
\rowcolor[gray]{0.95} \multicolumn{5}{c}{\textbf{Qwen2.5-Omni-7B}} \\
SnapKV    & 64.7 & 76.0 & 39.6 & 54.5 \\
AdaKV      & 64.7 & 76.0 & 40.8 & 54.9 \\
PyramidKV & 65.2 & 76.7 & 39.8 & 55.1 \\
AudioKV    & \textbf{66.7} & \textbf{77.3} & \textbf{42.2} & \textbf{55.5} \\
\bottomrule
\end{tabular}
\end{table}

\subsection{Generalization across AQA Benchmarks}

\label{sec:results-aqa}
We evaluate the generalization of various KV cache compression methods across four AQA datasets using Qwen2.5-Omni (3B and 7B), including \textbf{speech-chatbot-alpaca-eval} (S-Alpaca), \textbf{llama-questions} (Llama-Q), \textbf{speech-triviaqa} (S-Trivia), and \textbf{speech-web-questions} (S-Web). As shown in Table \ref{tab:aqa-generalization}, our proposed AudioKV consistently outperforms existing methods, such as SnapKV, AdaKV, and PyramidKV, across all benchmarks. 

Specifically, on the Speech-TriviaQA dataset, AudioKV achieves a significant improvement of approximately 1.3\% to 4.5\% over other baselines. The performance gain is consistent across both model scales, demonstrating that AudioKV more effectively preserves critical acoustic information while maintaining a compact KV cache.

\begin{figure}[ht]
  \centering
  \includegraphics[width=\linewidth]{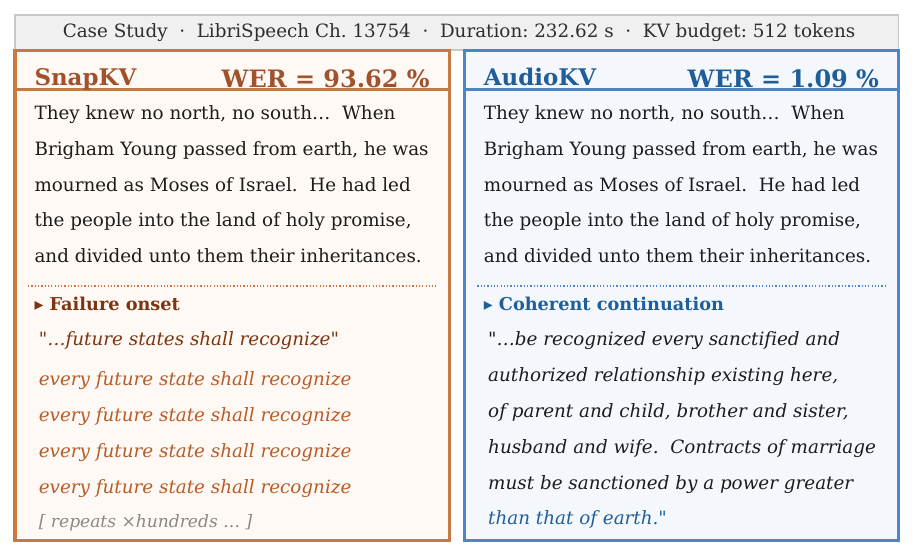}
  \caption{Qualitative visualization of long-form decoding behavior. SnapKV falls into a repetition loop, while AudioKV remains coherent and completes the passage.}
  \label{fig:failure-case-visualization}
\end{figure}

\subsection{Case Studies}

This case study analyzes a representative long-form ASR failure beyond average WER. As shown in Figure~\ref{fig:failure-case-visualization}, SnapKV enters a self-reinforcing repetition loop after a localized phrase transition. This suggests that its compressed KV cache loses sufficient global context to recover from a near-degenerate next-token distribution. Although the preceding transcription remains reasonable, the loop dominates subsequent generation and sharply increases WER.

In contrast, AudioKV maintains a coherent continuation through the same semantic region and completes the paragraph without collapse. This mode-level difference indicates that frequency-domain-aware retention better preserves discourse-level constraints over long horizons.

The example also explains the zero-valued results in our main tables. Under KV cache compression, repetition loops can cause the insertion count ($I$) in $\mathrm{WER}=(S+D+I)/N$ to grow far beyond the reference length ($N$). Evaluation scripts may therefore record the resulting divergent output as zero. Such values indicate catastrophic generation collapse rather than merely reduced transcription accuracy.

\subsection{Parameter Sensitivity Analysis of the Spectral Score Smoothing Component}
\begin{figure}[t]
    \centering 
    \includegraphics[width=\linewidth]{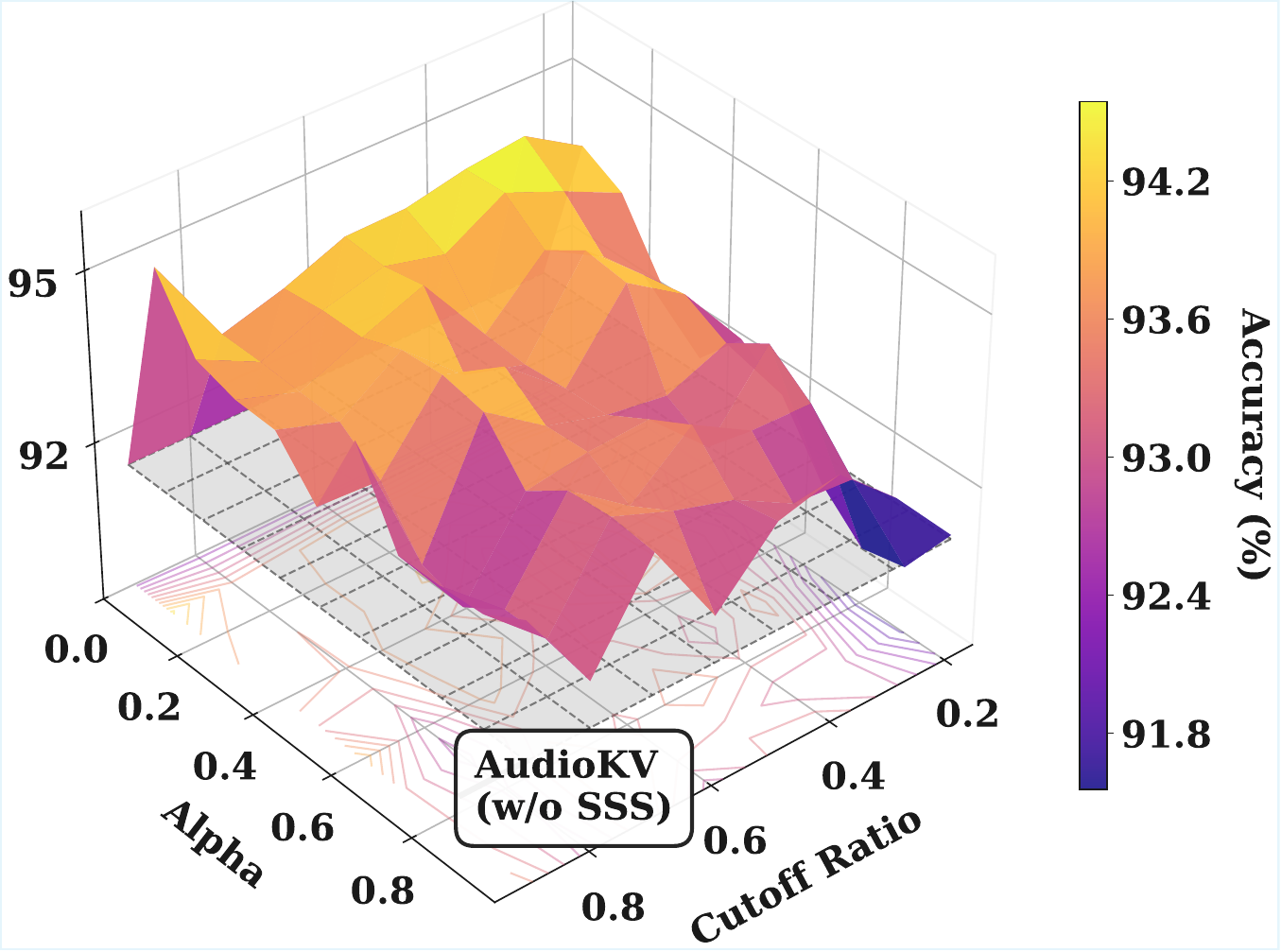}
    \caption{\textbf{Sensitivity analysis of SSS} on LibriSpeech-long with 50\% KV retention. The gray plane denotes the AudioKV baseline (91.4\%), recovered when $\alpha=0$. SSS remains robust across parameter settings and achieves up to 94.7\% accuracy.}
    \label{fig:fft_sensitivity}
\end{figure}

We evaluate the robustness of Spectral Score Smoothing (SSS) on LibriSpeech-long with a 50\% KV cache budget (Figure~\ref{fig:fft_sensitivity}). The gray plane denotes the AudioKV baseline (91.4\% accuracy), to which our method degenerates when $\alpha=0$. When SSS is activated ($\alpha > 0$), it consistently outperforms the baseline across a wide parameter space. Notably, the performance remains stable for $k^{\ast} \in [0.2, 0.8]$ and various $\alpha$. This stability is due to SSS treating importance scores as continuous signals and suppressing high-frequency noise via global frequency-domain filtering. The optimal configuration achieves 94.7\% accuracy, confirming that SSS is both effective and robust to hyperparameter choices.

\subsection{Efficiency Analysis}

\begin{figure}[t]
    \centering 
    \includegraphics[width=\linewidth]{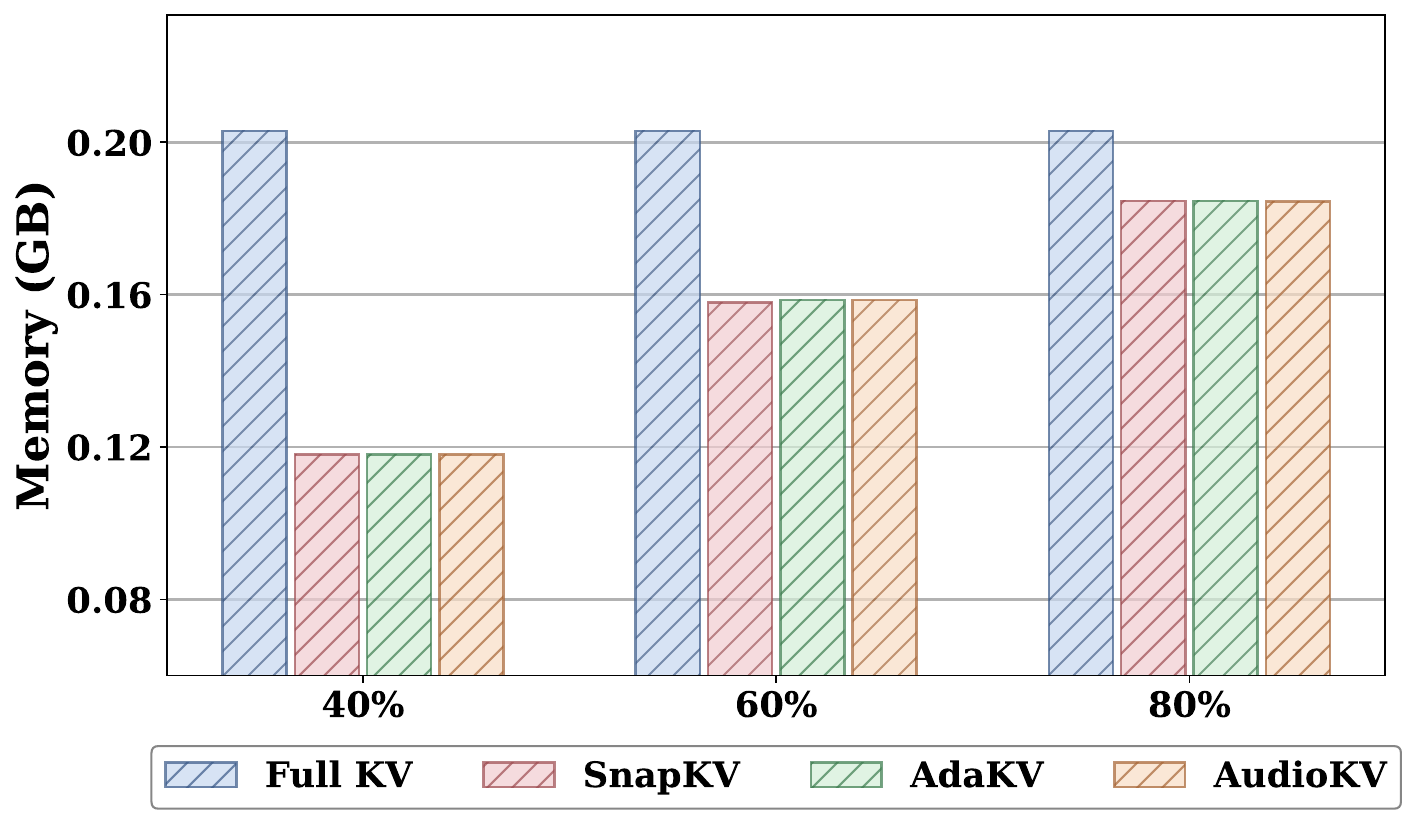}
    \caption{KV-cache memory footprint at retention ratios $r \in \{0.4, 0.6, 0.8\}$. Under identical memory budgets, AudioKV outperforms SnapKV and AdaKV, demonstrating higher utilization efficiency.}
    \label{fig:memory}
\end{figure}
We evaluate inference efficiency from the perspective of KV cache memory footprint. Figure~\ref{fig:memory} presents cache size under different compression ratios. Compared to the full KV cache, all compression-based methods substantially reduce memory consumption, confirming the effectiveness of KV compression in improving inference efficiency. Notably, although SnapKV, AdaKV, and AudioKV achieve similar cache sizes at the same compression ratio, our method consistently yields significantly superior task performance across all evaluated audio benchmarks under an identical memory budget. 

\label{sec:discussion}

\label{app:longform-failure:figure}

\section{Conclusion}
\label{sec:conclusion}
We proposed \textbf{AudioKV}, a KV cache compression framework tailored to head heterogeneity and acoustic continuity in LALMs. AudioKV combines \textbf{audio-aware head allocation} with \textbf{Spectral Score Smoothing (SSS)} to preserve speech-critical information and stabilize importance estimation. Experiments on Qwen and Gemma models outperform existing methods under tight memory budgets. At 40\% KV retention, AudioKV limits the accuracy drop of Qwen3-Omni-30B to 0.45\%, highlighting its effectiveness for memory-efficient long-context audio inference.

\begin{acks}
This work was supported by the National Natural Science Foundation of China (Grant No.62506318); Guangdong Provincial Department of Education Project (Grant No.2024KQNCX028); Scientific Research Projects for the Higher-educational Institutions (Grant No.2024312096), Education Bureau of Guangzhou Municipality; Guangzhou-HKUST(GZ) Joint Funding Program (Grant No.2025A03J3957), Education Bureau of Guangzhou Municipality.
\end{acks}

\clearpage
\bibliographystyle{ACM-Reference-Format}
\balance
\bibliography{AudioKV}

\end{document}